\documentclass[aps,showpacsx,twocolumn]{revtex4-1}
\usepackage{amssymb, amsmath}
\usepackage{mathbbold}
\usepackage{slashed}
\usepackage{graphicx}
\usepackage{dcolumn}
\usepackage{bm}
\usepackage{hyperref}
\usepackage{color}

\begin{document}

\title{ Topological Susceptibility in Three-Flavor Quark Meson Model at Finite Temperature}
\author{Yin Jiang$^{1,2}$}
\author{Tao Xia$^{1,3}$}
\author{Pengfei Zhuang$^1$}
\affiliation{$^1$Physics Department, Tsinghua University and Collaborative Innovation Center of Quantum Matter, Beijing 100084, China\\
             $^2$Physics Department and Center for Exploration of Energy and Matter, Indiana University, 2401 N Milo B. Sampson Lane, Bloomington, Indiana 47408, USA\\
             $^3$College of OptoElectronic Science and Engineering, National University of Defense Technology, Changsha 410073, China }
\date{\today}

\begin{abstract}
We study $U_A(1)$ symmetry and its relation to chiral symmetry at finite temperature through the application of functional renormalization group to the $SU(3)$ quark meson model.
Very different from the mass gap and mixing angel between $\eta$ and $\eta'$ mesons which are defined at mean field level and behavior like the chiral condensates, the topological susceptibility includes a fluctuations induced part which becomes dominant at high temperature. As a result, the $U_A(1)$ symmetry is still considerably broken in the chiral symmetry restoration phase.
\end{abstract}

\maketitle

\section{Introduction}
\label{s1}

It is well-known that the $U_A(1)$ symmetry is broken in the vacuum of Quantum Chromodynamics (QCD) by the anomaly due to the nontrivial topology of
the principal bundle of gauge field~\cite{thooft,leutwyler}, which leads to the nondegeneracy of $\eta$ and $\eta'$ mesons~\cite{witten,veneziano,rosenzweig}. As a strong interacting
system should approach its classic limit at high temperature,
all the broken symmetries including the $U_A(1)$ are expected to be restored in hot
medium~\cite{schafer}. While the relation between the $U_A(1)$ symmetry and chiral symmetry in
vacuum and at finite temperature has been studied for a
long time~\cite{kawarabayashi1,kawarabayashi2,alkofer}, it is still an open
question whether the $U_A(1)$ symmetry is restored in chiral symmetric phase.

The lattice simulation is a powerful tool to study
QCD symmetries. By a proper definition, the topological
charge and its susceptibility are used to describe the
$U_A(1)$ anomaly in the pure gauge field theory and the
unquenched theory~\cite{alles1,alles2}. In both cases, the susceptibility
drops down above the critical temperature $T_c$ of the
chiral restoration, but the charge keeps an obvious deviation
from zero at high temperature $T>T_c$. The simulation
for the instanton model shows such a partial restoration,
too~\cite{wanza}. From the recent lattice simulations of HotQCD~\cite{hotqcd} and JLQCD~\cite{jlqcd} collaborations, while the $U_A(1)$ symmetry is still broken at $T_c$ from both groups, the JLQCD claimed the $U_A(1)$ restoration at $T=1.2T_c$ and the HotQCD observed opposite result.

To clearly understand the relation between the $U_A(1)$ and chiral symmetries, we need to put the QCD system in chiral limit where the chiral phase transition at high temperature is well defined. In real case with nonzero quark mass, the chiral symmetry cannot be fully restored by thermodynamics, and therefore one possible mechanism of the $U_A(1)$ breaking at high temperature is the residual chiral breaking. Since the chiral limit cannot be realized in lattice calculations where a nonzero pion mass is always used, we need to consider effective models to clarify the relation between the two symmetries. Two of the often employed models are the Nambu--Jona-Lasinio (NJL)~\cite{nambu} model at quark level~\cite{vogl,klevansky,volkov,hatsuda,buballa} and the linear sigma model at hadron level~\cite{gellmann, cjt} and including quarks (quark-meson model)~\cite{gasiorrowicz, grid}. At finite temperature and density, the two models are widely used to discuss chiral and $U_A(1)$ properties of strongly interacting matters, see for instance \cite{zhuang,bilic,schaffner,fukushima,toublan,barducci,ebert,mao,abuki,andersen,schaefer}.

In this work we use the functional renormalization
group (FRG) method to study $U_A(1)$ symmetry and its relation to chiral symmetry in the quark-meson model.
As a nonperturbative method, the FRG~\cite{frgreview1, frgreview2} has been used to study phase transitions in various systems
like cold atom gas~\cite{floerchinger}, nucleon gas~\cite{friman}, and hadron gas~\cite{herbst,stokic,bohr,blaizot,schaefer2,fukushima2}. By solving the flow equation which connects
physics at different momentum scales, the FRG shows a
great power to describe the phase transitions and the corresponding
critical phenomena which are normally difficult
to be controlled in the mean-field approximation
because of the absence of quantum fluctuations. Instead
of adding hot loops to the thermodynamic potential in the
usual ways of going beyond mean field, the fluctuations are included in the FRG effective
action through running the RG scale from ultraviolet limit to infrared limit, which, as an advantage, can automatically guarantee the
Nambu-Goldstone theorem in the symmetry breaking
phase. Based on our previous works in the linear sigma~\cite{jiang} and NJL~\cite{xia} models where we focused on the meson masses, we calculate here in the $SU(3)$ quark-meson model the topological susceptibility and $\eta-\eta'$ mixing angle which describe directly and clearly the degree of $U_A(1)$ symmetry breaking. We will see that while the
$U_A(1)$ is controlled by chiral condensates in the chiral breaking phase, it is dominated by fluctuations
after the chiral symmetry is restored.

The paper is organized as follows. In Section \ref{s2} we first define in the quark-meson model the correspondent of the topological charge density $Q$ of QCD and then calculate analytically the topological susceptibility $\chi$. In Section \ref{s3} we briefly review the FRG application to the quark-meson model and introduce the pseudoscalar meson's mixing angle $\theta_P$. In Section \ref{s4} we numerically solve the FRG flow equations with grid method and show the temperature dependence of the scalar and pseudoscalar mesons as well as the mixing angle and the topological susceptibility. We summarize in Section \ref{s5}.

\section{Topological Susceptibility in $SU(3)$ Quark Meson Model}
\label{s2}

The topological susceptibility is a fundamental correlation
function in QCD and is the key to understanding
the dynamics in the $U_A(1)$ channel. In
this section we calculate the topological susceptibility
at finite temperature within the framework
of the three-flavor quark-meson model.

In QCD the axial current $J_5^\mu=\bar\psi\gamma^\mu\gamma_5\psi$ is not conserved due to the $U_A(1)$ anomaly induced by the instanton effect,
\begin{equation}
\label{j5}
\partial_\mu J_5^\mu=2N_f Q(x)+2im_0\bar\psi \gamma_5\psi,
\end{equation}
where $m_0$ is the current quark mass, $N_f=3$ the number of flavors, and $Q$ the topological charge density
\begin{equation}
\label{qx}
Q(x)={g^2\over 32\pi^2}F^a_{\mu\nu}\tilde F_a^{\mu\nu}
\end{equation}
with the gluon field strength tensor $F^a_{\mu\nu}$ and the coupling constant $g$ between quark and gluon fields. The topological susceptibility $\chi$ is defined as the Fourier
transform of the connected correlation function $\langle T\left(Q(x)Q(0)\right)\rangle$,
\begin{equation}
\label{chi}
\chi=\int d^4x \langle T\left(Q(x)Q(0)\right)\rangle_{\rm connected},
\end{equation}
where $T$ denotes the time-ordering operator.

We now define the correspondent of the topological charge density $Q$ in the three-flavor quark-meson model through the conservation law (\ref{j5}). The Lagrangian density of the model contains the meson section and quark section,
\begin{equation}
\label{l1}
{\cal L}={\cal L}_m+{\cal L}_q,
\end{equation}
the coupling between quark and meson fields is included in the quark section. Taking renormalizablity into account in Minkowski space the meson section ${\cal L}_m$ reads
\begin{eqnarray}
{\cal L}_m &=& {\rm Tr}\left[\partial_\mu\Phi\partial^\mu\Phi^\dagger\right]-\left(m^2\rho_1+\lambda_1\rho^2_1+\lambda_2\rho_2\right)\nonumber\\
&&+c\xi+{\rm Tr}\left[H(\Phi+\Phi^\dagger)\right],
\end{eqnarray}
where the meson matrix $\Phi = T^a\phi_a$ and the trace Tr are
defined in the flavor space, the meson fields $\phi_a=\sigma_a+i\pi_a$ contains 9 scalar mesons $\sigma_a$ and 9 pseudoscalar mesons $\pi_a$,
the $3\times 3$ Gell-Mann matrices $T_a=\lambda_a/2$ for $a=1,\cdots,8$ and $T_0=\mathbf{1}/\sqrt 6$ for $a=0$ obey the relations ${\rm Tr}\left(T_aT_b\right)=\delta_{ab}/2$, $\left[T_a,\ T_b\right]=if_{abc}T_c$ and $\{T_a,\ T_b\}=d_{abc}T_c$ with the
structure constants $f_{abc}$ and $d_{abc}$, $m^2$ is the meson mass parameter,
$c, \lambda_1$ and $\lambda_2$ are the couplings among mesons, and $\rho_i$ for $i=1,2$ are the chiral symmetry invariants $\rho_i={\rm Tr}\left(\Phi\Phi^\dagger\right)^i$.

The $U_A(1)$ symmetry breaking is through the term $c\xi$ with $\xi=\det \Phi+\det\Phi^\dagger$ which mimics the $U_A(1)$ anomaly of QCD. Note that the kinetic term ${\rm Tr}[\partial_\mu\Phi\partial^\mu\Phi^\dagger]$ and the $U_A(1)$ breaking term preserves the $SU_L(3)\times SU_R(3)$ chiral symmetry.

The quark section ${\cal L}_q$ reads
\begin{equation}
{\cal L}_q=\bar \psi(i\gamma^\mu\partial_\mu-m_0+\mu\gamma^0-g\Phi_5)\psi,
\end{equation}
where the quark-meson interaction is through the meson matrix $\Phi_5=T^a(\sigma_a+i\gamma_5\pi_a)$ with the coupling constant $g$. Since we focus on the temperature behavior of the $U_A(1)$ and chiral symmetries in this work, we neglect in the following the quark chemical potential matrix $\mu$.

The terms ${\rm Tr}\left[H(\Phi+\Phi^\dagger)\right]$ in ${\cal L}_m$ and $m_0\bar\psi\psi$ in ${\cal L}_q$ break explicitly the chiral symmetry of the system and lead to nonzero pion mass in vacuum, where the matrix $H$ is defined as $H=h_aT_a$ with 9 parameters $h_a$.

For the $U_A(1)$ transformation at quark level,
\begin{equation}
\psi \rightarrow e^{-i\theta_A\gamma_5T^0}\psi
\end{equation}
with the QCD vacuum angle $\theta_A$, or
\begin{equation}
\label{tq}
\psi\rightarrow \psi-i\theta_A\gamma_5 \psi/\sqrt 6
\end{equation}
for an infinite small transformation, one has accordingly the transformation for mesons in the quark-meson model
\begin{eqnarray}
\bar\psi_m\psi_n &\rightarrow& \bar\psi_m\psi_n-2\theta_A\bar\psi_mi\gamma_5\psi_n/\sqrt 6,\nonumber\\
\bar\psi_mi\gamma_5\psi_n &\rightarrow& \bar\psi_mi\gamma_5\psi_n+2\theta_A\bar\psi_m\psi_n/\sqrt 6,
\end{eqnarray}
or
\begin{eqnarray}
\sigma_a &\rightarrow& \sigma_a-2\theta_A\pi_a/\sqrt 6, \nonumber\\
\pi_a &\rightarrow& \pi_a+2\theta_A\sigma_a/\sqrt 6
\end{eqnarray}
which can be expressed in a compact way,
\begin{eqnarray}
\label{tm}
\Phi &\rightarrow& \left(1+i 2\theta_A/\sqrt 6\right)\Phi,\nonumber\\
\det\Phi &\rightarrow& \left(1+i\sqrt 6\theta_A\right)\det\Phi.
\end{eqnarray}

Under the transformations (\ref{tq}) and (\ref{tm}), only the Kobayashi-Maskawa-'t Hooft (KMT) term $c\xi$ and the two explicit chiral breaking terms $m_0\bar\psi\psi$ and ${\rm Tr}\left[H(\Phi+\Phi^\dagger)\right]$ in the Lagrangian density (\ref{l1}) change with the variation
\begin{eqnarray}
\label{dl1}
\Delta {\cal L} &=& i2\theta_A/\sqrt 6\big[3c\left(\det\Phi-\det\Phi^\dagger\right)\nonumber\\
&&+m_0\bar\psi\gamma_5\psi+{\rm Tr}\left(H\left(\Phi-\Phi^\dagger\right)\right)\big].
\end{eqnarray}

On the other hand, according to Noether's theorem, the variation
of the Lagrangian density by the $U_A(1)$ transformation can be written as
\begin{eqnarray}
\label{dl2}
\Delta{\cal L}&=&\partial_\mu\frac{\partial{\cal L}}{\partial(\partial_\mu\psi)}\Delta\psi + \partial_\mu\frac{\partial{\cal L}}{\partial(\partial_\mu\sigma_a)}\Delta\sigma_a+ \partial_\mu\frac{\partial{\cal L}}{\partial(\partial_\mu\pi_a)}\Delta\pi_a\nonumber\\
&=&\partial_\mu\left[\bar\psi i\gamma^\mu\Delta\psi+\partial^\mu\sigma_a\Delta\sigma_a+\partial^\mu\pi_a\Delta\pi_a\right]\nonumber\\
&=&\partial_\mu\left[{\theta_A\over \sqrt 6}\left(\bar\psi \gamma^\mu\gamma_5\psi-2\partial^\mu\sigma_a\pi_a
+2\partial^\mu\pi_a\sigma_a\right)\right],
\end{eqnarray}
where we have used the explicit expression of the meson kinetic term
\begin{equation}
{\rm Tr}\left[\partial_\mu\Phi\partial^\mu\Phi^\dagger\right]={1\over 2}\left(\partial_\mu\sigma_a\partial^\mu\sigma_a+\partial_\mu\pi_a\partial^\mu\pi_a\right).
\end{equation}

From the comparison of (\ref{dl1}) with (\ref{dl2}), we have the conservation law in the quark-meson model,
\begin{eqnarray}
\partial_\mu J_5^\mu &=& \partial_\mu\left(\bar\psi\gamma^\mu\gamma_5\psi-
2\partial^\mu\sigma_a\pi_a+2\partial^\mu\pi_a\sigma_a\right)\\
&=&-12c{\rm Im}\det\Phi+2im_0\bar\psi\gamma_5\psi+2i{\rm Tr}\left(H\left(\Phi-\Phi^\dagger\right)\right).\nonumber
\end{eqnarray}
Taking the definition of the topological charge density (\ref{j5}) and considering the meson degrees of freedom in the quark-meson model, the above conservation law defines the charge density $Q(x)$ in the model,
\begin{equation}
Q(x)=-2c{\rm Im}\det\Phi(x),
\end{equation}
which is purely induced by the KMT term.

With the $18$ scalar and pseudoscalar mesons, the charge density can be explicitly expressed as a sum of all possible products of three meson fields,
\begin{eqnarray}
\label{q}
Q&=&{c\over 2}\Bigg[\sqrt{2\over 27} \pi _0^3-{1\over \sqrt{27}} \pi _8^3-{1\over \sqrt 6}\pi_0\left(\sum_{a=1}^8\left(\pi _a^2-\sigma_a^2\right)+2 \sigma _0^2\right)\nonumber\\
&&+{1\over 2}\pi _3\left(\sum_{a=4}^5\left(\pi _a^2-\sigma_a^2\right)-\sum_{a=6}^7\left(\pi _a^2-\sigma_a^2\right)\right)\nonumber\\
&&+{1\over \sqrt 3}\pi _8\left(\sum_{a=1}^3\left(\pi _a^2-\sigma_a^2\right)-{1\over 2}\sum_{a=4}^7\left(\pi _a^2-\sigma _a^2\right)+\sigma_8^2\right)\nonumber\\
&&+\pi _1\left(\sum_{a=4}^5\left(\pi_a \pi_{a+2}-\sigma _a \sigma _{a+2}\right)+\sqrt {2\over 3} \sigma _0 \sigma _1-{2\over \sqrt 3}\sigma _1 \sigma _8\right)\nonumber\\
&&+\pi _2\Bigg(\pi _5 \pi _6-\pi _4 \pi _7+\sqrt {2\over 3}\sigma _0 \sigma _2-\sigma _5 \sigma _6+\sigma _4\sigma _7\nonumber\\
&&-{2\over \sqrt 3}\sigma _2 \sigma _8\Bigg)+\sqrt{2\over 3} \pi _3\left(\sigma _0 \sigma _3-\sqrt{2}\sigma _3 \sigma _8\right)\nonumber\\
&&+\pi _4\left(\sqrt{2\over 3}\sigma _0 \sigma _4-\sigma _3 \sigma _4-\sigma _1 \sigma _6+\sigma _2 \sigma _7+{1\over \sqrt 3}\sigma _4 \sigma _8\right)\nonumber\\
&&+\pi_5\left(\sqrt{2\over 3}\sigma _0 \sigma _5-\sigma _3 \sigma _5-\sigma _2 \sigma _6-\sigma _1 \sigma _7+{1\over \sqrt 3}\sigma _5 \sigma _8\right)\nonumber\\
&&-\pi _6\left(\sigma _1 \sigma _4-\sqrt{2\over 3}\sigma _0 \sigma_6-\sigma _3 \sigma _6+\sigma _2 \sigma _5-{1\over\sqrt 3}\sigma _6 \sigma _8\right)\nonumber\\
&&+\pi _7\left(\sigma _2 \sigma _4+\sqrt{2\over 3}\sigma _0 \sigma _7-\sigma _1 \sigma _5+\sigma _3 \sigma _7+{1\over \sqrt 3}\sigma _7 \sigma _8\right)\nonumber\\
&&+\sqrt{2\over 3} \pi _8 \sigma _0 \sigma _8\Bigg].
\end{eqnarray}

\begin{figure}[!hbt]
\begin{center}
\includegraphics[width=0.4\textwidth]{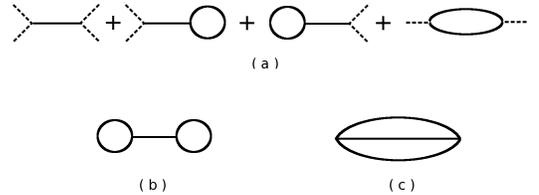}
\caption{ Diagrammatic representation of the topological susceptibility $\chi$ with (a) and without (b, c) explicit condensate contribution. The dashed and solid lines indicate respectively the meson condensates and propagators.}
\label{fig1}
\end{center}
\end{figure}

Having obtained the expression of the topological
charge density $Q(x)$ in the quark-meson model, we calculate the topological susceptibility $\chi$ according to the
definition (\ref{chi}). By separating the fields $\varphi_i=\sigma_a,\pi_a$ into the condensate and fluctuation parts $\varphi_i(x)=\langle\varphi_i\rangle+\varphi_i^{\rm fl}(x)$ and following Wick$'$s theorem, we take a full contraction in the connected correlation function $\langle T\left(Q(x)Q(0)\right)\rangle$ in
terms of the meson condensates $\langle\varphi_i\rangle$ and the meson propagators $G_{ij}(x,y)=G_{ii}(x,y)\delta_{ij}=\langle\varphi_i^{\rm fl}(x)\varphi_i^{\rm fl}(y)\rangle\delta_{ij}$,
\begin{eqnarray}
\label{chi2}
\chi&=&\left({c\over 12\sqrt 6}\right)^2\sum_{i,j,k,l,m,n=\sigma_a,\pi_a}\int d^4x\Big[\nonumber\\
&&a_{ijklmn}\langle\varphi_i\rangle\langle\varphi_j\rangle G_{kl}(x,0)\langle\varphi_m\rangle\langle\varphi_n\rangle\nonumber\\
&&+ b_{ijklmn}\langle\varphi_i\rangle\langle\varphi_j\rangle G_{kl}(x,0)G_{mn}(0,0)\nonumber\\
&&+ c_{ijklmn}G_{ij}(x,x)G_{kl}(x,0)\langle\varphi_m\rangle\langle\varphi_n\rangle\nonumber\\
&&+ d_{ijklmn}\langle\varphi_i\rangle G_{jm}(x,0)G_{kl}(x,0)\langle\varphi_n\rangle\nonumber\\
&&+ e_{ijklmn}G_{ij}(x,x)G_{kl}(x,0)G_{mn}(0,0)\nonumber\\
&&+ f_{ijklmn}G_{in}(x,0) G_{jm}(x,0)G_{kl}(x,0)\Big],
\end{eqnarray}
where the terms without the propagator $G(x,0)$ between the two points $x$ and $0$ are excluded from the connected correlation function. The first four terms in the square bracket which are all with condensates are diagrammatically shown in Fig.\ref{fig1}a, and the fifth and sixth terms which contain only closed propagators $G(x,x)$ and $G(0,0)$ and propagators $G(x,0)$ between the space-time points $x$ and $0$ are shown in Fig.\ref{fig1}b and Fig.\ref{fig1}c. Since only the scalar mesons $\sigma_0$ and $\sigma_8$ can couple to the vacuum
without violating Lorentz invariance and parity, the classical field $\langle\varphi\rangle$ contains only two components $\langle\sigma_0\rangle$ and $\langle\sigma_8\rangle$. This largely reduces the terms in (\ref{chi2}) and simplifies the calculation of $\chi$.

It is clear that the four terms shown in Fig.\ref{fig1}a control the susceptibility $\chi$ in the chiral breaking phase at low temperature where the condensates are nonzero. However, the last two terms shown in Fig.\ref{fig1}b and Fig.\ref{fig1}c become dominant in the symmetry restoration phase at high temperature where the condensates vanish in chiral limit and fluctuations characterize the system. Note that, the $U_A(1)$ symmetry breaking results in off-diagonal propagators $G_{08}(x,y)$ and $G_{80}(x,y)$, but by diagonalizing the subspace $a=0,8$ the off-diagonal elements disappear and there exist only diagonal propagators $G_{ij}(x,y)=G_{ii}(x,y)\delta_{ij}$. The lowest order contribution to the correlation comes from the first diagram in Fig.\ref{fig1}a which involves four condensates and one propagator $G(x,0)$,
\begin{eqnarray}
\label{0pi}
\int d^4x G_{ll}(x,0) &=& {1\over (2\pi)^4}\int d^4x d^4p G_{ll}(p)e^{ip\cdot x}\nonumber\\
&=&{1\over M_l^2}.
\end{eqnarray}
This term governs the topological susceptibility before
the chiral restoration and the temperature dependence is from the condensates $\langle\sigma_0\rangle(T)$ and $\langle\sigma_8\rangle(T)$ and mass $M_l(T)$ for the meson species $l$, which will be calculated in the framework of functional renormalization group in next section.

For the closed propagators $G(x,x)$ and $G(0,0)$, shown as 1PI diagrams in Fig.\ref{fig1}a and Fig.\ref{fig1}b, by doing the Matsubara frequency summation in the imaginary time formalism of finite temperature field theory, one has
\begin{eqnarray}
\label{1loop}
G_{ll}(x,x) &=& G_{ll}(0,0)\nonumber\\
&=&\int {d^4p\over (2\pi)^4} {i\over p^2-M_l^2}\nonumber\\
&=& T\int{d^3{\bf p}\over (2\pi)^3}\sum_n{1\over \omega_n^2+\epsilon_l^2}\nonumber\\
&=& \int {d^3{\bf p}\over (2\pi)^3}{f(\epsilon_l)\over \epsilon_l},
\end{eqnarray}
where $\omega_n=2n\pi T$ with $n=0,\pm 1,\pm2,\cdots$ are the Boson frequencies, $f(\epsilon_l)=1/\left(e^{\beta\epsilon_l}-1\right)$ is the Bose-Einstein distribution function with $\beta=1/T$ and meson energy $\epsilon_l=\sqrt{{\bf p}^2+M_l^2}$, and we have subtracted the divergent term $1/\epsilon_l$ in the last step by a simple renormalization.

The last diagram in Fig.\ref{fig1}a includes a 2PI loop between the two condensates,
\begin{eqnarray}
\label{2pi}
&&\int d^4x G_{ll}(x,0)G_{mm}(x,0)\nonumber\\
&=& \int {d^4xd^4pd^4q\over (2\pi)^8} G_{ll}(p)G_{mm}(q)e^{i(p+q)x}\nonumber\\
&=& T^2\int{d^3{\bf p} d^3{\bf q}\over (2\pi)^3} \sum_{j,k}{\beta\delta_{jk}\delta({\bf p+q})\over (\omega_j^2+\epsilon_l^2)(\omega_k^2+\epsilon_k^2)}\nonumber\\
&=& T^2\int{d^3{\bf p} d^3{\bf q}\over (2\pi)^3} \sum_{j,k}{\left(e^{i\beta\omega_j}-e^{i\beta\omega_k}\right)\delta({\bf p+q})\over (i\omega_j-i\omega_k) (\omega_j^2+\epsilon_l^2)(\omega_k^2+\epsilon_m^2)}\nonumber\\
&=& \int {d^3{\bf p}d^3{\bf q}\over (2\pi)^3}{\delta({\bf p+q})\over \epsilon_l^2-\epsilon_m^2}\left({f(\epsilon_l)\over \epsilon_l}-{f(\epsilon_m)\over \epsilon_m}\right)
\end{eqnarray}
with the meson energies $\epsilon_l=\sqrt{{\bf p}^2+M_l^2}$ and $\epsilon_m=\sqrt{{\bf q}^2+M_m^2}$, where we have subtracted again the divergent terms in the last step.

Like Fig.\ref{fig1}b, Fig.\ref{fig1}c comes purely from the quantum fluctuations and does not depend on the chiral condensate explicitly. With similar technique for the Matsubara frequency summation used in (\ref{2pi}), we have
\begin{eqnarray}
\label{2pi2}
&&\int d^4x G_{ll}(x,0)G_{mm}(x,0)G_{nn}(x,0)\nonumber\\
&=& \int {d^4xd^4pd^4qd^4k\over (2\pi)^{12}} G_{ll}(p)G_{mm}(q)G_{nn}(k)e^{i(p+q+k)x}\nonumber\\
&=& T^3\int{d^3{\bf p} d^3{\bf q} d^3{\bf k}\over (2\pi)^6} \sum_{i,j,k} {\beta\delta_{i,j+k} \delta({\bf p+q+k})\over (\omega_i^2+\epsilon_l^2)(\omega_j^2+\epsilon_m^2)(\omega_k^2+\epsilon_n^2)}\nonumber\\
&=& \int {d^3{\bf p} d^3{\bf q} d^3{\bf k}\over (2\pi)^6}\delta({\bf p+q+k})\sum_{ijk=lnm,mln,nml}\Bigg[\nonumber\\
&&{\epsilon_i^2-\epsilon_j^2-\epsilon_k^2\over \left(\epsilon_n^2-\epsilon_l^2-\epsilon_m^2\right)^2-4\epsilon_l^2\epsilon_m^2}{f(\epsilon_j)f(\epsilon_k)\over \epsilon_j\epsilon_k}\nonumber\\
&&+{\epsilon_i+\epsilon_j\over \left(\epsilon_i+\epsilon_j\right)^2-\epsilon_k^2}{f(\epsilon_k)\over 2\epsilon_l\epsilon_m\epsilon_n}\Bigg]
\end{eqnarray}
with the meson energies $\epsilon_l=\sqrt{{\bf p}^2+M_l^2},\ \epsilon_m=\sqrt{{\bf q}^2+M_m^2}$ and $\epsilon_n=\sqrt{{\bf k}^2+M_n^2}$, where we have used the relationship between two distribution functions $f(\epsilon_l)f(\epsilon_m)/f(\epsilon_l+\epsilon_m)=f(\epsilon_l)+f(\epsilon_m)+1$ and again subtracted the divergent terms not accompanied by any distribution function.

The momentum integration of the three terms with two distribution functions in (\ref{2pi2}) is convergent obviously and can further be simplified. For instance, it reads
\begin{eqnarray}
&&\int {d^3{\bf p} ^3{\bf q} d^3{\bf k}\over (2\pi)^6}{\delta({\bf p+q+k})\left(\epsilon_n^2-\epsilon_l^2-\epsilon_m^2\right)\over \left(\epsilon_n^2-\epsilon_l^2-\epsilon_m^2\right)^2-4\epsilon_l^2\epsilon_m^2}{f(\epsilon_l)f(\epsilon_m)\over \epsilon_l\epsilon_m}\nonumber\\
=&&\int  {dp dq pq\over 32\pi^4}\ln\left|{\left(\left(\epsilon_l+\epsilon_m\right)^2-\epsilon_n^{+2}\right)\left(\left(\epsilon_l-\epsilon_m\right)^2-\epsilon_n^{+2}\right)\over \left(\left(\epsilon_l+\epsilon_m\right)^2-\epsilon_n^{-2}\right)\left(\left(\epsilon_l-\epsilon_m\right)^2-\epsilon_n^{-2}\right)}\right|\nonumber\\
&&\times {f(\epsilon_l)f(\epsilon_m)\over \epsilon_l \epsilon_m}
\end{eqnarray}
with the meson energies $\epsilon_n^+=\sqrt{({\bf p+q})^2+M_n^2}$ and $\epsilon_n^-=\sqrt{({\bf p-q})^2+M_n^2}$. The momentum integration of the other three terms with only one distribution function in (\ref{2pi2}) is divergent and the renormalization is done in Ref.~\cite{baacke}. For instance, it reads
\begin{eqnarray}
&&\int {d^3{\bf p} d^3{\bf q} d^3{\bf k}\over (2\pi)^6} \delta({\bf p+q+k})
{\epsilon_m+\epsilon_n\over \left(\epsilon_m+\epsilon_n\right)^2-\epsilon_l^2}{f(\epsilon_l)\over 2\epsilon_l\epsilon_m\epsilon_n}\nonumber\\
=&&{1\over 32\pi^4}\Bigg[-\int_0^1d\alpha\ln\left(\alpha{M_m^2\over M_l^2}+(1-\alpha)\left({M_n^2\over M_l^2}-\alpha\right)\right)\nonumber\\
&&-\gamma_E+\ln\left(4\pi{\mu^2\over M_l^2}\right)\Bigg]\int dp p^2{f(\epsilon_l)\over \epsilon_l}
\end{eqnarray}
with the Euler constant $\gamma_E$ and the factorization scale $\mu=1$ GeV.

\section{Quantization with Functional Renormalization Group}
\label{s3}

We now review the application of the functional renormalization group to the SU(3) quark-meson model, the details can be seen in Refs.\cite{cjt,schaffner,grid,schaefer,jiang}. The core quantity in the
framework of FRG is the averaged effective action $\Gamma_k$ at a momentum scale $k$ in Euclidean space. In quantum field theory, fluctuations are included in the effective action $\Gamma$
by functionally integrating the classical action,
\begin{equation}
\Gamma\left[\Phi,\psi\right]=\int \mathcal{D}\Phi^\dagger\ \mathcal{D}\Phi\ \mathcal{D}\bar\psi \mathcal{D}\psi e^{-S_{cl}\left[\Phi, \psi\right]}
\end{equation}
with $S_{cl}\left[\Phi, \psi\right] = \int d^4x\ \mathcal{L}(\Phi, \psi)$. However, working out this integration is almost impossible, if there is any interaction involving in the Lagrangian density.
As an effective way adopted in FRG, an averaged action which is a function of the renormalization group scale $k$ is introduced~\cite{frgreview1},
\begin{equation}
\Gamma_k\left[\Phi, \psi\right]=\int \mathcal{D}\Phi^\dagger\ \mathcal{D}\Phi\ \mathcal{D}\bar\psi \mathcal{D}\psi e^{-\left(S\left[\Phi, \psi\right]+\Delta S_k\left[\Phi, \psi\right]\right)},
\end{equation}
where the scale dependence is carried by the additional action $\Delta S_k\left[\Phi, \psi\right]=\int d^4 x\left[\rm{Tr}\left(\Phi^\dagger R_k^B\Phi\right)+\bar\psi R_k^F\psi\right]$, and the infrared cutoff functions $R_k^B$ for bosons and $R_k^F$ for fermions should be properly chosen to suppress the fluctuations at low momentum. Once $k$ approaches to zero, there would be no fluctuations suppressed. In this
way, all the fluctuations are gradually included as $k$ evolves from the ultraviolet limit to the infrared limit. Details of the evolution is coded in the flow equation for the averaged action $\Gamma_k$~\cite{frgreview1},
\begin{eqnarray}
\label{weteq}
k\partial_k\Gamma_k=\frac{1}{2}{\rm Tr}
\frac{k\partial_k R_k^B}{\Gamma_k^{B(2)}+R_k^B}-{\rm Tr}
\frac{k\partial_k R_k^F}{\Gamma_k^{F(2)}+R_k^F},
\end{eqnarray}
where $\Gamma_k^{B(2)}$ and $\Gamma_k^{F(2)}$ are second order functional derivatives of $\Gamma_k$
with respect to the boson and fermion fields. In our calculation below, we choose the cutoff functions as the optimized regulators~\cite{frgreview2}
\begin{equation}
R_k^B({\bf p})={\bf p}^2\left(\frac{k^2}{{\bf p}^2}-1\right)\Theta\left(1-\frac{{\bf p}^2}{k^2}\right)
\end{equation}
for bosons and
\begin{equation}
R_k^F({\bf p})=i \slashed {\bf p}\left(\sqrt{\frac{k^2}{{\bf p}^2}}-1\right)\Theta\left(1-\frac{{\bf p}^2}{k^2}\right)
\end{equation}
for fermions.

To solve the flow equation, we take the local potential approximation~\cite{frgreview1} which is good enough if we consider only the chiral condensates and meson spectra. At this level, all the fluctuations are supposed to be included in an effective potential $U_k\left(\Phi\right)$ which is reduced to the classical potential
\begin{eqnarray}
U_\Lambda(\langle\Phi\rangle)&=&m^2\langle\rho_1\rangle+\lambda_{1}\langle\rho_1\rangle^2+\lambda_{2}\langle\rho_2\rangle\nonumber\\
&&-c\langle\xi\rangle-h_0\langle\sigma_0\rangle-h_8\langle\sigma_8\rangle
\end{eqnarray}
at the ultraviolet limit $k=\Lambda$ where all the fluctuations are supposed to vanish. Assuming homogeneous condensates and after doing directly the momentum integration and Matsubara frequency summation at finite temperature, the flow equation (\ref{weteq}) is simplified as a partial differential equation for the effective potential~\cite{jiang},
\begin{equation}
\label{floweq}
\partial_k U_k(\langle\Phi\rangle)=\frac{k^4}{12\pi^2}
\left[ \sum_{b}\frac{1}{E_b}\coth\frac{E_b}{2T}-12\sum_{f}\frac{1}{E_f}\tanh\frac{E_f}{2T}\right]
\end{equation}
with 18 boson and 3 fermion energies $E_b=\sqrt{k^2+M_b^2}, (b=\pi_a,\sigma_a)$ and $E_f=\sqrt {k^2+M_f^2}, (f=u,d,s)$.

The two independent condensates $\langle\sigma_0\rangle$ and $\langle\sigma_8\rangle$ or the light and strange quark condensates $\langle\sigma_u\rangle=\sqrt{2\over 3}\langle\sigma_0\rangle+\sqrt{1\over 3}\langle\sigma_8\rangle$ and $\langle\sigma_s\rangle=\sqrt{1\over 3}\langle\sigma_0\rangle-\sqrt{2\over 3}\langle\sigma_8\rangle$ are determined by the minimization of the potential,
\begin{equation}
\label{gap}
{\partial U_k(\langle\Phi\rangle)\over \partial\langle\sigma_u\rangle}={\partial U_k(\langle\Phi\rangle)\over \partial\langle\sigma_s\rangle}=0.
\end{equation}
This leads to the scale dependence of the condensates $\langle\sigma_u\rangle_k$ and $\langle\sigma_s\rangle_k$. The dynamical quark and meson masses are defined as the coefficients of the quadratic terms $\bar\psi\psi$, $\pi_a^\text{fl}\pi_b^\text{fl}$ and $\sigma_a^\text{fl}\sigma_b^\text{fl}$ in the lagrangian density after the separations $\pi_a=\langle\pi_a\rangle+\pi_a^\text{fl}$ and $\sigma_a=\langle\sigma_a\rangle+\sigma_a^\text{fl}$,
\begin{eqnarray}
M_u &=& M_d=m_0+{1\over 2}g\langle\sigma_u\rangle_k,\nonumber\\
M_s &=& m_0+{1\over \sqrt 2}g\langle\sigma_s\rangle_k,\nonumber\\
\left(M_S^2\right)_{ab} &=& {\partial^2 U_k(\Phi)\over \partial\sigma_a\partial\sigma_b}\Big|_{\Phi\to\langle\Phi\rangle},\nonumber\\
\left(M_P^2\right)_{ab} &=& {\partial^2 U_k(\Phi)\over \partial\pi_a\partial\pi_b}\Big|_{\Phi\to\langle\Phi\rangle}.
\end{eqnarray}
The meson masses are just the eigenvalues of the curvature of the effective potential $U_k(\Phi)$. They form two $9\times 9$ matrices $M_S^2$ and $M_P^2$, and $7$ of their diagonal elements are the masses of the scalar mesons $a_0$ and $\kappa$ and pseudoscalar mesons $\pi$ and $K$.
Due to the $U_A(1)$ breaking, there exists one independent nonzero off-diagonal element for each matrix, $\left(M_S^2\right)_{08}=\left(M_S^2\right)_{80}$ and $\left(M_P^2\right)_{08}=\left(M_P^2\right)_{80}$. Diagonalizing the meson subspace $a=0,8$ generates the pseudoscalar mesons $\eta$ and $\eta'$ and the corresponding scalar mesons which are the eigen states of the Hamiltonian of the model,
\begin{eqnarray}
\eta_0=\cos{\theta_P}\eta-\sin{\theta_P}\eta'\nonumber\\
\eta_8=\sin{\theta_P}\eta+\cos{\theta_P}\eta',
\end{eqnarray}
where $\theta_P$ is the mixing angle in the pseudoscalar channel and can be expressed in terms of the masses,
\begin{equation}
\tan 2\theta_P = {2\left(M_P^2\right)_{08}\over \left(M_P^2\right)_{00}-\left(M_P^2\right)_{88}}.
\end{equation}
In chiral limit, it is reduced to
\begin{equation}
\tan 2\theta_P=2\sqrt 2
\end{equation}
in the chiral restoration phase, which leads to a constant mixing angle $\theta\simeq 35^0$ after the phase transition. In real case, we can expand the angle in terms of the chiral condensate $\langle\sigma_u\rangle$ at high temperature where chiral symmetry is partially restored and $\langle\sigma_u\rangle$ becomes small,
\begin{equation}
\tan 2\theta_P=2\sqrt 2\left(1-{9\langle\sigma_u\rangle\over 2\left(\left(M_S^2\right)_{11}-\left(M_S^2\right)_{44}\right)}\right)+{\mathcal O}\left(\langle\sigma_u\rangle^2\right).
\end{equation}
For the scalar channel, we can introduce the mixing angle $\theta_S$ in a similar way. In chiral limit, there is no mixing $\theta_S=0$ in the chiral symmetry restoration phase due to $\left(M_S^2\right)_{08}=0$.

It is necessary to note that we can analytically prove the Goldstone theorem corresponding to the spontaneous chiral symmetry breaking in the FRG frame.

\section{Numerical Results}
\label{s4}

We now numerically solve the flow equation (\ref{floweq}) for the effective potential $U_k$ together with the gap equations for the condensates $\langle\sigma_u\rangle$ and $\langle\sigma_s\rangle$. The both sides of the flow equation depend only on $\langle\sigma_u\rangle$ and $\langle\sigma_s\rangle$ or $\langle\rho_1\rangle$ and $\langle\rho_2\rangle$, it is then a first order differential equation with initial condition $U_{k=\Lambda}$ at the ultraviolet limit, and we can numerically solve the effective potential as a whole in a two-dimensional grid~\cite{grid}. The evolution of the potential is evaluated by discretizing the potential in the plane of $\langle\rho_1\rangle$ and $\langle\rho_2\rangle$. We also adopt the clamped cubic splines to evaluate the derivatives of the potential with respect to $\langle\rho_1\rangle$ and $\langle\rho_2\rangle$ and interpolate the potential in order to find the global minimum.

We first solve the flow equation in vacuum. We choose the ultraviolet momentum $\Lambda=1$ GeV which is the typical scale of effective models at hadron level. The initial potential $U_\Lambda$ is so chosen to fit the pseudoscalar meson masses $M_\pi$, $M_K$, $M_\eta$ and $M_{\eta'}$, decay constants $f_\pi$ and $f_K$, and dynamical quark mass or chiral condensate $\langle\sigma_u\rangle$ in vacuum. For our calculation in real case, we take the renormalization parameters $m_\Lambda^2=$ (867.76 MeV)$^2$, $\lambda_{1\Lambda}=-32/3$ and $\lambda_{2\Lambda}=50$, the $U_A(1)$ breaking parameter $c=4807.84$ MeV, the chiral breaking parameters $h_u$=2(120.73 MeV)$^2$ and $h_s=\sqrt 2$ (336.41 MeV)$^2$, and the Yukawa coupling strength $g=6.5$.
Considering the fact that the system at high enough momentum
is dominated by the dynamics and not affected remarkably by the temperature, the temperature dependence
of the initial condition of the flow equation at the ultraviolet momentum
can be safely neglected. Therefore, we take the temperature
independent initial condition $U_\Lambda(T)=U_\Lambda$ in vacuum.

\begin{figure}
\includegraphics[scale=0.4]{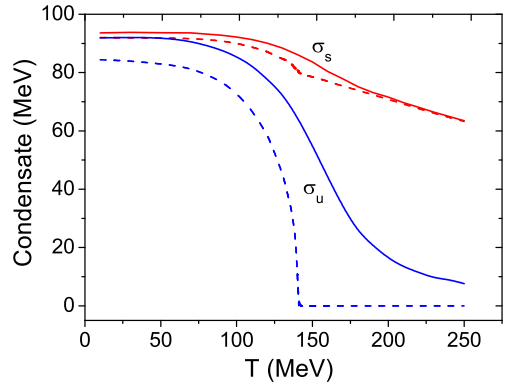}
\caption{(Color online) The light and strange quark condensates $\langle\sigma_u\rangle/2$ and $\langle\sigma_s\rangle/\sqrt 2$ as
functions of temperature in chiral limit (dashed lines) and real case (solid lines).}
\label{fig2}
\end{figure}
We now show the temperature dependence of the chiral condensates in Fig.\ref{fig2}. In chiral limit with $h_u=0$, the light quark condensate $\langle\sigma_u\rangle$ which is the order parameter of the chiral phase transition continuously drops down with temperature and goes to zero at the critical temperature $T_c=140$ MeV. In real case with nonzero $h_u$, the chiral phase transition becomes a crossover, and the order parameter decreases very rapidly around the critical temperature. The temperature dependence of the strange quark condensate $\langle\sigma_s\rangle$ is rather smooth in comparison with the light quark condensate, it decreases with temperature gradually and is nonzero in the symmetry restoration phase.

\begin{figure}
\includegraphics[scale=0.5]{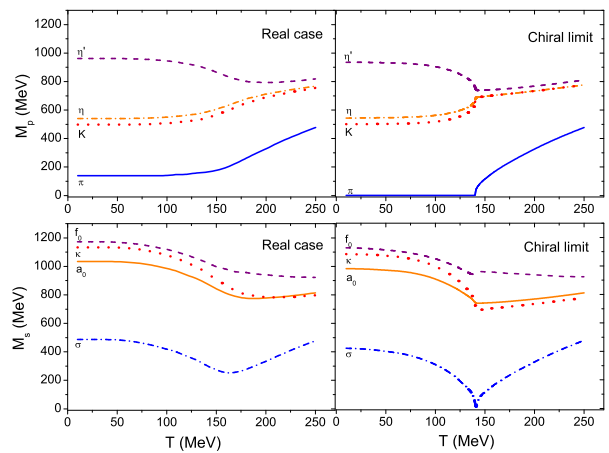}
\caption{(Color online) The scalar and pseudoscalar meson masses $M_S$ and $M_P$ as functions of temperature in chiral limit (right panel) and real case (left panel).}
\label{fig3}
\end{figure}
The masses of the 9 scalar mesons $\sigma, \kappa, f_0$ and $a_0$ and 9 pseudoscalar mesons $\pi, K, \eta$ and $\eta'$ are shown in Fig.\ref{fig3} as functions of temperature. In chiral limit, $\pi$s are the three Goldstone modes, and their mass keeps zero in the chiral breaking phase. At the critical point, $\sigma$ becomes also massless. In the symmetry restoration phase, it is easy to find $\left(M_S^2\right)_{00}=\left(M_P^2\right)_{11}$ and $\left(M_S^2\right)_{44}=\left(M_P^2\right)_{44}$ which mean the degeneration of $\pi$s and $\sigma$ and $K$s and $\kappa$s. In real case, the degeneration disappears, but the corresponding scalar and pseudoscalar mesons approach to each other at high temperature. While the $\eta$ and $\eta'$ mass splitting becomes much weaker in the chiral restoration phase than in the symmetry breaking phase, it does not vanish. This indicates $U_A(1)$ symmetry breaking even at extremely high temperature.

\begin{figure}
\includegraphics[scale=0.4]{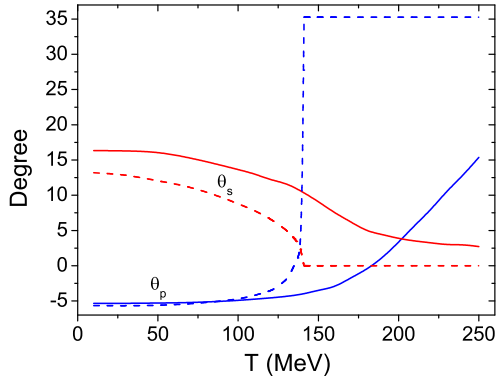}
\caption{(Color online) The pseudoscalar and scalar mixing angles $\theta_P$ and $\theta_S$ as functions of temperature in chiral limit (dashed lines) and real case (solid lines).}
\label{fig4}
\end{figure}
Fig.\ref{fig4} shows the mixing angles $\theta_P$ and $\theta_S$ as functions of temperature in chiral limit and real case. At $T=0$ the angles are determined by the meson masses in vacuum. In chiral limit, the pseudoscalar angle increases with temperature from the starting value $\theta_P\simeq -5^0$, then crosses zero and jumps up suddenly at the critical point of chiral phase transition, and finally keeps as a constant $\theta_P\simeq 35^0$ in the symmetry restoration phase, as we analyzed in the last section. In real case, the sudden jump disappears and the angle gradually approaches to $35^0$ in high temperature limit. The temperature behavior of the scalar angle $\theta_S$ is very similar to the chiral condensate $\langle\sigma_u\rangle$. In contrast with $\theta_P$, it drops down continuously with increasing temperature. In the chiral restoration phase it disappears in chiral limit and is still sizeable in real case.

\begin{figure}[!hbt]
\begin{center}
\includegraphics[width=0.4\textwidth]{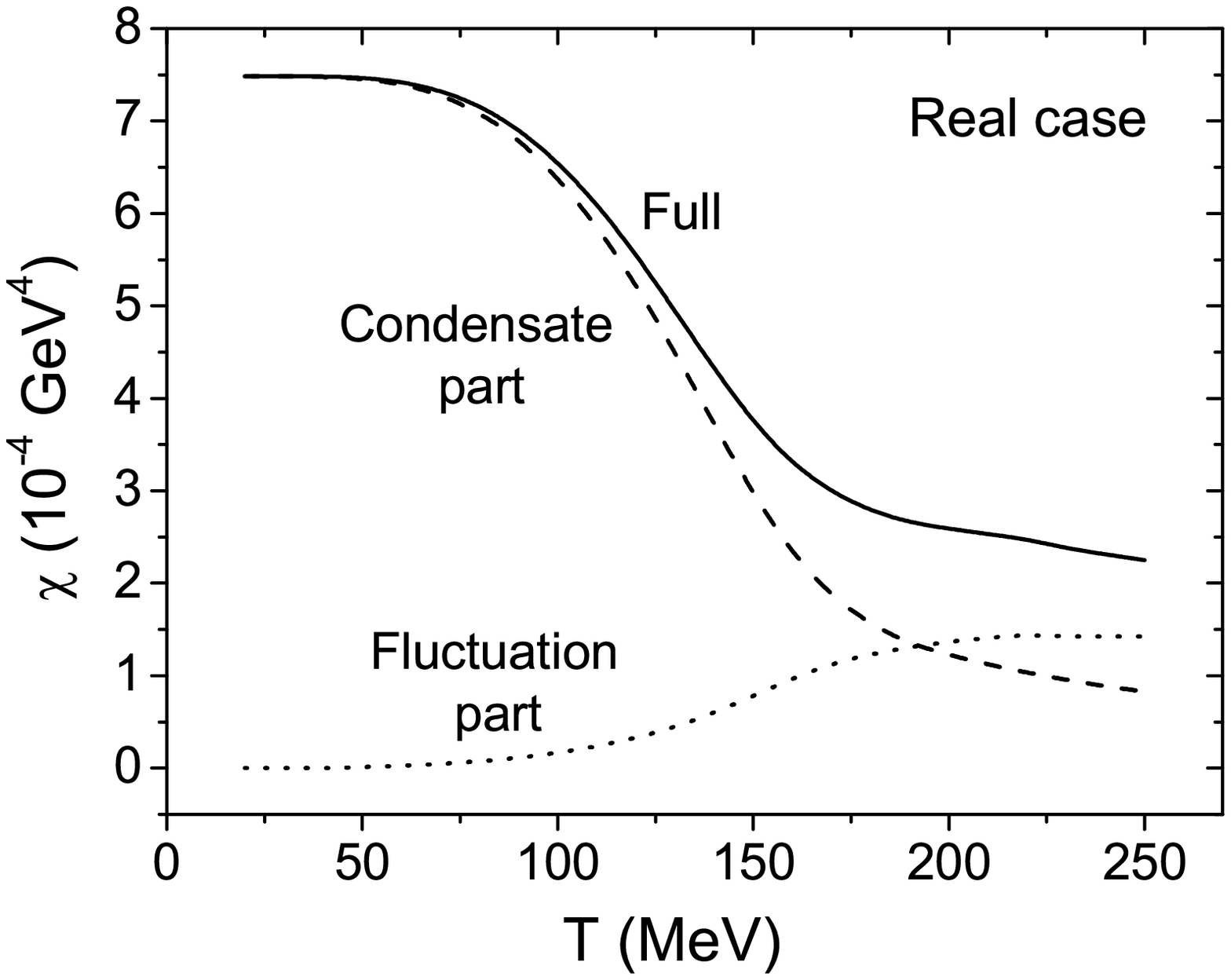}
\includegraphics[width=0.4\textwidth]{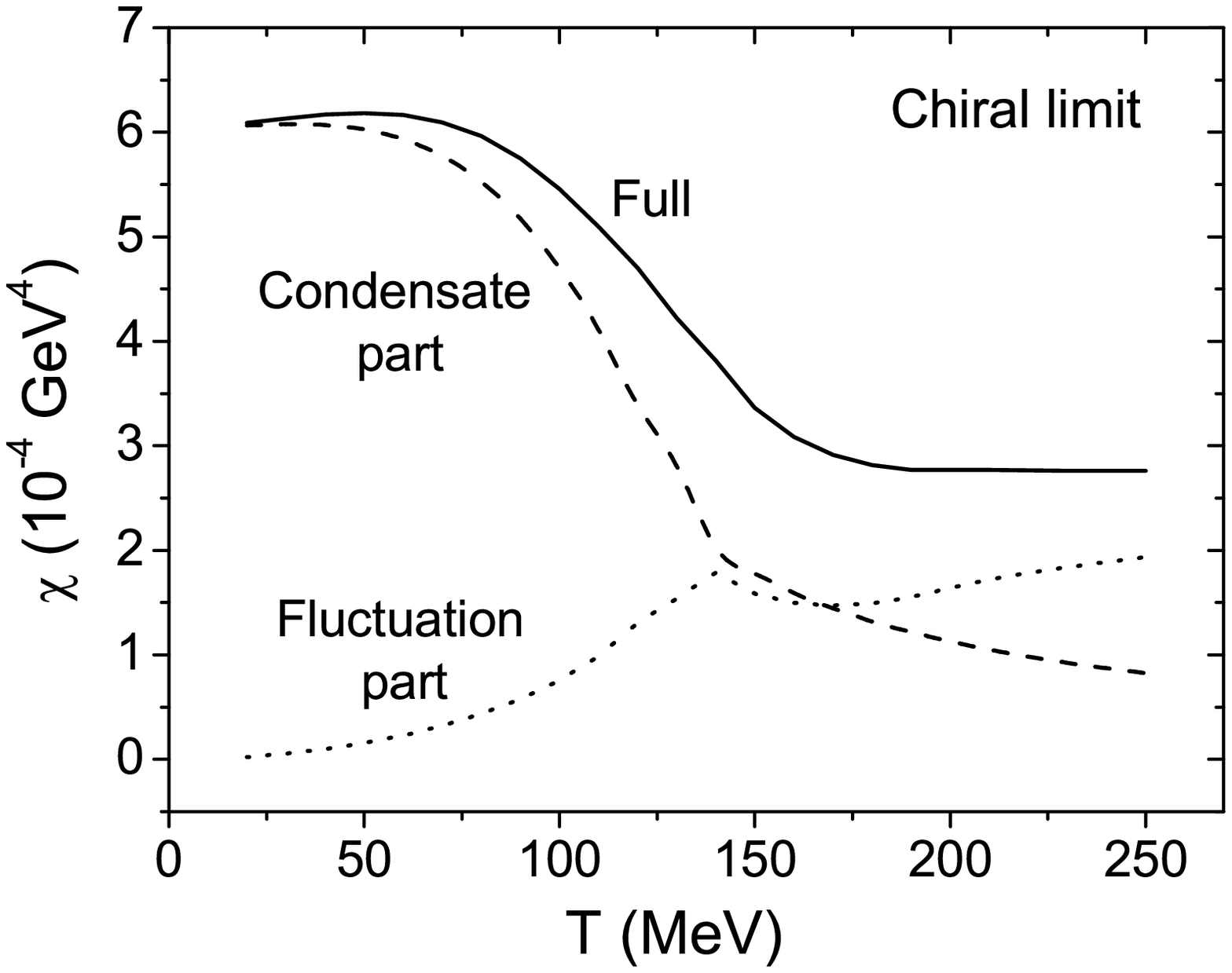}
\caption{The topological susceptibility $\chi$ as a function of temperature in chiral limit (left panel) and real case (right panel). Dashed and dotted lines are the contributions controlled by the condensates and fluctuations, respectively, and solid lines are the full results. }
\label{fig5}
\end{center}
\end{figure}
Now we come to the topological susceptibility $\chi$ which is the most straightforward criterion for the quantum anomaly. From its expression shown in Eq.(\ref{chi2}) or Fig.\ref{fig1}, it contains the condensates dominated part and the fluctuations induced part, indicated respectively by dashed and dotted lines in Fig.\ref{fig5}. Using the known condensates $\langle\sigma_u\rangle$ and $\langle\sigma_s\rangle$ and the meson mass matrices  $M_S^2$ and $M_P^2$ with off-diagonal elements $\left(M_S^2\right)_{08},  \left(M_S^2\right)_{80}, \left(M_P^2\right)_{08}$ and $\left(M_P^2\right)_{80}$, we can directly calculate the susceptibility by summarizing all the 6-field correlations in (\ref{chi2}). An alternative way is to diagonalizing the subspace with $a=0,8$ and use the 18 scalar and pseudoscalar eigen states of the model and the mixing angels $\theta_S$ and $\theta_P$. The two ways of calculations are equivalent. In both cases of chiral limit and real case, while the condensates control the susceptibility in the chiral symmetry breaking phase and around the critical point, the fluctuations become the dominant contribution at high temperature. Different from the condensates controlled part which drops down continuously with increasing temperature, the fluctuations induced part goes up with temperature. As a result, there will be still $U_A(1)$ symmetry breaking in the chiral symmetry restoration phase.

\section{Conclusion}
\label{s5}
We investigated the $U_A(1)$ symmetry and its relation to the chiral symmetry at finite temperature, by applying the functional renormalization group to the $SU(3)$ quark meson model. We calculated the mass gap and mixing angel between $\eta$ and $\eta'$ mesons and the topological susceptibility to see if the $U_A(1)$ symmetry is restored at high temperature. Since the mass gap and mixing angle are defined through meson masses at mean field level, the former approaches to zero and the latter becomes a constant in the chiral symmetry  restoration phase. This means that the two symmetries are restored at almost the same critical temperature. However, this is the conclusion in mean field approximation. When the fluctuations are included in the calculation of the topological susceptibility which is the most straightforward criterion for the anomaly, the conclusion is very different. The susceptibility contains two parts, the condensates controlled part which behaviors like the mass gap and mixing angle and the fluctuations induced part which becomes dominant in the chiral restoration phase. The former drops down and the latter goes up with increasing temperature. As a result of the competition, the full susceptibility is still remarkably large when the chiral symmetry is restored.
It is necessary to note that our discussion here on $U_A(1)$ and chiral symmetries is in the scope of hadrons. Beyond the scope, the quantum and thermal fluctuations at meson level will breakdown due to meson melting in hot medium, and the susceptibility should finally disappear at extremely high temperature because of the non-trivial gluon configurations fading out.

\begin{acknowledgments}
The work is supported by the NSFC and MOST grant Nos. 11335005, 11575093, 2013CB922000 and 2014CB845400.
\end{acknowledgments}

\end{document}